\documentclass[prl,aps,twocolumn,superscriptaddress]{revtex4-1}
\usepackage{graphicx,color}
\usepackage{amsthm}
\usepackage{amsfonts}
\usepackage{algorithmic}
\usepackage{enumerate}
\usepackage{latexsym}
\usepackage{amsmath}
\usepackage{amssymb}
\usepackage[colorlinks=true,citecolor=blue,linkcolor=blue]{hyperref}

\usepackage{dcolumn}
\usepackage{bm}

\usepackage[T1]{fontenc}
\usepackage{mathptmx}

\begin{document}
\title{The competition between antiferromagnetism and superconductivity in a doped Hubbard model with anisotropic interaction}
\author{Runyu Ma}
\affiliation{Department of Physics, Beijing Normal University, Beijing 100875, China}
\affiliation{Beijing Computational Science Research Center, Beijing 100193, China}
\author{Tianxing Ma}
\email{txma@bnu.edu.cn}
\affiliation{Department of Physics, Beijing Normal University, Beijing 100875, China}

\begin{abstract}
The competition between antiferromagnetism and superconductivity is one of the central questions in the research
of strong correlated systems. In this work, we utilize a double layer model containing Hubbard interaction
and interlayer Heisenberg interaction to reveal their competitions. This model is free of sign problem at certain
conditions, and we perform projector quantum Monte Carlo simulations to extract the ground state correlations of
magnetism and superconductivity. Our results shows that the superconductivity emerges when the antiferromagnetism is suppressed
by tuning the filling or the anisotropy of the interlayer Heisenberg interaction. This model can be seen as an analogue of unconventional superconductors and may help us to understand the transition from an antiferromagnetic insulator to
a superconductor.
\end{abstract}

\maketitle
\section{Introduction}
It is widely known that the unconventional superconductors may have magnetic parents\cite{RevModPhys.66.763}.
Since the discovery of superconductivity in doped cuprates\cite{Bednorz1986},
large number of experiments have been conducted to explore its mechanism
and complex phase diagram\cite{RevModPhys.84.1383,RevModPhys.87.855,RevModPhys.93.025006},
especially the transition from an antiferromagnetic (AFM) insulator to
a superconductor, which is a critical part in the study of unconventional superconductivity.
Theoretically, constructing a model that can describe these phenomenon especially the
competition between antiferromagnetism and superconductivity is an import problem for condense matter physicists.
Hubbard-like model, has been proven to be a good candidate to describe AFM insulators\cite{PhysRevB.80.075116, PhysRevLett.62.591,PhysRevLett.124.017003} and superconductors\cite{PhysRevLett.63.1996,Tranquada1995,RevModPhys.84.1383, PhysRevB.84.180513}.
However, because of its exponential growth of Hilbert space,
the introduction of strong electronic interaction brings new challenges into solving this model.

Many numerical methods have been developed to solve Hubbard model and its extensions.
For examples, density matrix renormalization group\cite{RevModPhys.77.259},
quantum Monte Carlo\cite{ACIOLI199775}, dynamic mean field theory\cite{RevModPhys.78.865} and so on.
Among them, quantum Monte Carlo is a great method for its advantage of accuracy and
the convenience of use in some sense, which has been used to extract ground state or finite
temperature properties of strong correlated system.
In past decades, quantum Monte Carlo simulations of Hubbad-like model have achieved fruitful results, including pairing symmetries\cite{JPSJ.76.113708,PhysRevB.105.245131,PhysRevB.103.235156,PhysRevB.103.144514},
charge density wave state\cite{PhysRevB.101.205139}, localization of
electronic states\cite{PhysRevLett.120.116601,PhysRevB.104.045116,PhysRevLett.100.076602},
unconventional superconductivity in twisted bilayer graphene\cite{PhysRevB.105.L121110, PhysRevLett.126.027002,PhysRevB.98.121406,j.scib.2019.01.026}, stripe order in two-dimensional electronic correlated system\cite{science.aam7127,PhysRevX.10.031016}.

Analogue to the cuprate superconductors, we are particularly interesting in the doping case where superconductivity emerges.
However, quantum Monte Carlo algorithms are limited by the sign problem, especially when we want to use them to investigate the transition from an AFM insulator to a superconductor.
At finite doping  where superconductivity emerges, the sign problem is severe in the original Hubbard model and undermine the accuracy of simulations. There are some works attempt to eliminate or alleviate sign problem. For examples, expressing spinless fermion in a Majorana representation can make the simulations avoid sign problem\cite{PhysRevB.91.241117}, adiabatically switching on the electronic interaction\cite{PhysRevLett.127.217003}, constraining the phase space\cite{PhysRevB.55.7464}. There are also some works attempt to utilize sign problem to analyze quantum critical points\cite{science.abg9299}. Besides, the sign problem can be eliminated by some special symmetries.
For example, 
bipartite lattice like square or honeycomb lattice can avoid sign problem at half filling because of the particle-hole symmetry.
Another example is the attractive Hubbard model, it can avoid sign problem at arbitrary filling, because after HS transformation in charge channel, spin up is identical to spin down, and so its determinant is positive definite.

Recently, a sign free extended bilayer Hubbard-like model, an generalized Scalapino-Zhang-Hanke model, is utilized to investigate the transition from AFM insulating state to superconducting (SC) state\cite{PhysRevB.106.054510}, which  provide an excellent platform to study the unconventional superconductivity. Through time reverse symmetry, this effective model is free of the sign problem at arbitrary filling. It is shown that a quantum
phase transition occurs from an Ising anisotropic AFM insulating
phase or an SU(2) invariant Mott insulating phase without
the AFM ordering to a rung-singlet SC phase with an extended
$s$-wave symmetry driven by doping. This is an attractive feature to conduct quantum Monte Carlo simulations at
finite doping. However, in that work, the parameters are confined in a relative small region at some fixed terms, to establish AFM long range order at half filling.

The interlayer or inter-orbital interactions may be an import part in some
certain materials, like monolayer FeSe\cite{PhysRevB.94.155127}.  This kind of interaction, maybe also implemented in ultracold atom experiments\cite{Demler2022}, which could provide a possible
platform to observe such competition between antiferromagnetism and superconductivity. Thus in this article, we further conduct a more comprehensive investigation on this extended bilayer Hubbard model and its interlayer interactions, and extract its ground state properties by using projector quantum Monte Carlo (PQMC) algorithm.
We have tuned the strength and the anisotropy of interlayer Heisenberg interaction, and we find some interesting behaviours which is not revealed in the previous work. By calculating the correlation lengths and using the finite size scaling technique, we carefully investigate the magnetism and superconductivity in this model. We find that the system is sensitive to both the doping and anisotropy of interlayer interaction. The antiferromagnetism will fade away when hole doping is introduced, and then superconductivity appears.
At the same doping, the anisotropy of interlayer interaction also affects the superconductivity and magnetism,
where the $J_{\bot}$ part of Heisenberg interaction may be a critical component when SC pairs are taking shape.
In some sense, this interlayer Heisenberg interaction is similar to $t$-$J$ model, which can be considered as a good starting point to study cuprate superconductors\cite{RevModPhys.78.17}. The $t$-$J$ model contains Heisenberg interactions at the nearest neighbor.
This is different from our model where only interlayer interactions are possessed, and then we have different pairing symmetries.
However, the mechanics behind them maybe the same. Our results reveal the competition between antiferromagnetism and superconductivity in a numerical exact manner, which is import for us to understand the transition from an AFM insulator to a
superconductor.

\section{Model and Method}
The effective model we construct is on a two-layer square lattice, including a Hubbard interaction term and an anisotropic Heisenberg interaction term\cite{PhysRevB.71.155115,PhysRevLett.91.186402}. The Hamiltonian can be written as fellow,
\begin{equation}
\begin{aligned}
H= &-t \sum_{\langle {\bf i,j}\rangle, l \sigma}c^{\dagger}_{{\bf i}l\sigma} c_{{\bf j}l\sigma} \\
&+\sum_{{\bf i}}[g_1\left( c^{\dagger}_{{\bf i}1\uparrow}c_{{\bf i}1\uparrow}-c^{\dagger}_{{\bf i}1\downarrow}c_{1\downarrow}-c^{\dagger}_{{\bf i}2\uparrow}c_{{\bf i}2\uparrow}+c^{\dagger}_{{\bf i}2\downarrow}c_{{\bf i}2\downarrow}\right)^2 \\
&+g_2\left( c^{\dagger}_{{\bf i}1\uparrow}c_{{\bf i}1\downarrow} +c^{\dagger}_{{\bf i}1\downarrow}c_{{\bf i}1\uparrow} -c^{\dagger}_{{\bf i}2\uparrow}c_{{\bf i}2\downarrow} -c^{\dagger}_{{\bf i}2\downarrow}c_{{\bf i}2\uparrow}\right)^2\\
&+g_2\left( -i c^{\dagger}_{{\bf i}1\uparrow}c_{{\bf i}1\downarrow}+ ic^{\dagger}_{{\bf i}1\downarrow}c_{{\bf i}1\uparrow}+ ic^{\dagger}_{{\bf i}2\uparrow}c_{{\bf i}2\downarrow}- i c^{\dagger}_{{\bf i}2\downarrow}c_{{\bf i}2\uparrow}\right)^2]\\
=&-t \sum_{\langle {\bf i,j}\rangle, l \sigma}c^{\dagger}_{{\bf i}l\sigma} c_{{\bf j}l\sigma} +
 \sum_{{\bf i}} [J_{z} S^{z}_{{\bf i}1} S^{z}_{{\bf i}2}+\frac{1}{2} J_{\bot} \left(S^{+}_{{\bf i}1} S^{-}_{{\bf i}2}+S^{-}_{{\bf i}1} S^{+}_{{\bf i}2}\right)] \\
&+\sum_{{\bf i}}U \left(n_{{\bf i}1\uparrow} n_{{\bf i}1\downarrow} + n_{{\bf i}2\uparrow} n_{{\bf i}2\downarrow}\right)- \frac{U}{2} N
\end{aligned}
\label{eq:1}
\end{equation}
where we define $U=-2g_1 - 4g_2$, $J_z = -8g_1$, $J_{\bot} = -8g_2$ for notational convenience. In this equation, $c_{{\bf i}l\sigma}$($c^{\dagger}_{{\bf i}l\sigma}$) means annihilating(creating) an electron at site ${\bf i}$, layer $l=1,2$, spin $\sigma=\uparrow, \downarrow$ and $\langle {\bf i,j} \rangle$ indicates the nearest neighbor.
Besides, $S^{z}_{{\bf i}l} = \frac{1}{2}n_{{\bf i}l\uparrow}-\frac{1}{2}n_{{\bf i}l\downarrow}$, $S^{+/-}_{{\bf i}l} = c^{\dagger}_{{\bf i}l\uparrow/\downarrow}c_{{\bf i}l\downarrow/\uparrow}$ and $n_{{\bf i}l\sigma} = c^{\dagger}_{{\bf i}l\sigma}c_{{\bf i}l\sigma}$.

The main idea of the PQMC algorithm is applying projector to a trial wave function $\vert \Psi_T \rangle$, and the observables can be computed
\begin{equation}
\langle \hat{O} \rangle = \lim_{\beta \rightarrow \infty} \frac{\langle \Psi_{T} \vert e^{-\beta \hat{H}/2} \hat{O} e^{-\beta \hat{H}/2} \vert \Psi_{T} \rangle}{\langle \Psi_{T} \vert e^{-\beta \hat{H}} \vert \Psi_{T} \rangle}
\end{equation}
By using HS transformation, we can write
\begin{equation}
\frac{\langle \Psi_{T} \vert e^{-\beta \hat{H}/2} \hat{O} e^{-\beta \hat{H}/2} \vert \Psi_{T} \rangle}{\langle \Psi_{T} \vert e^{-\beta \hat{H}} \vert \Psi_{T} \rangle}
\approx \sum_{s} \mathbf{P}_s \langle \hat{O} \rangle_{s}
\end{equation}
where $\langle \hat{O} \rangle_{s} = \frac{\langle \Psi_{T} \vert e^{-\beta \hat{H}_{s}/2} \hat{O} e^{-\beta \hat{H}_{s}/2} \vert \Psi_{T} \rangle}{\langle \Psi_{T} \vert e^{-\beta \hat{H}_{s}} \vert \Psi_{T} \rangle}$
is the expectation at certain auxiliary field $s$, and $\mathbf{P}_s = \frac{det(P^{\dagger} e^{-\beta \hat{H}_{s}} P)}{\sum_{s} det(P^{\dagger} e^{-\beta \hat{H}_{s}} P)}$ ($P$ is the matrix form of $\vert \Psi_{T} \rangle$),
can be seen as the weight of Monte Carlo sampling. In general, the $\mathbf{P}_s$ is not
positive definite, then the sign problem occurs.

In this work, we resort to time-reverse symmetry to avoid the sign problem, which restrict the form of interaction.
If the action after HS transformation should possess time reverse symmetry\cite{PhysRevB.71.155115},
its eigenvalues are always complex conjugate pairs, this ensure the positive definite of its determinant.
As for Eq.~\ref{eq:1}, if $g_1$ and $g_2$ are both negative, the matrix after HS transformation $\hat{H}_{s}$ have time-reverse symmetry, and this can ensure the positive definite of $\mathbf{P}_{s}$. For more details, see Refs\cite{PhysRevB.71.155115,PhysRevLett.91.186402}.
In this work, we use noninteracting ground state wave function as trial wave function, and random
chemical potentials are added on every sites, otherwise the degeneracy shall break the wave function's symmetries and lead to sign problem.

Next, we define antiferromagnetic and superconducting order and their correlations,
\begin{equation}
S{\mathbf{k}} = \frac{1}{L^2} \sum_{\bf i, j} e^{-i \mathbf{k} \cdot (\mathbf{R}_{\bf i} - \mathbf{R}_{\bf j})} \langle (S^{z}_{{\bf i}1} - S^{z}_{{\bf i}2}) (S^{z}_{{\bf j}1} - S^{z}_{{\bf j}2}) \rangle
\end{equation}
where $L$ is lattice size, and we denote $S_{AFM} = S{(\pi, \pi)}$. The SC correlation is defined as
\begin{equation}
P{\mathbf{k}} = \frac{1}{L^2} \sum_{\bf i, j} e^{-i \mathbf{k} \cdot (\mathbf{R}_{\bf i} - \mathbf{R}_{\bf j})} \langle \Delta^{\dagger}_{\bf i} \Delta_{\bf j} \rangle
\end{equation}
where $\Delta_{\bf i} = c_{{\bf i}1\uparrow}c_{{\bf i}2\downarrow} - c_{{\bf i}1\downarrow}c_{{\bf i}2\uparrow} + c_{{\bf i}2\uparrow}c_{{\bf i}1\downarrow} - c_{{\bf i}2\downarrow}c_{{\bf i}1\uparrow}$, and we denote $P_{sc} = P{(0, 0)}$.

One of the major challenges is the expensive computational cost, and most of our results are confined to system size $L = 8$.
To characterize the competition quantitatively at a fixed system size, we resort to the correlation length defined as\cite{PhysRevB.91.165108,PhysRevB.89.094516}
\begin{equation}
\xi(L)^2 = \frac{1}{4\sin\left(\pi / L\right)^2}\left(\frac{C_{k}}{C_{k+\delta k}} - 1\right)
\end{equation}
where $\delta k$ is the minimum momentum of size $L$. The correlation length may not reflect the long range order accurately since our simulations are confined to a small lattice size, but it can reveal the competition between observables directly.

\section{Results and Discusses}
\begin{figure}[h]
\includegraphics[width=0.45\textwidth]{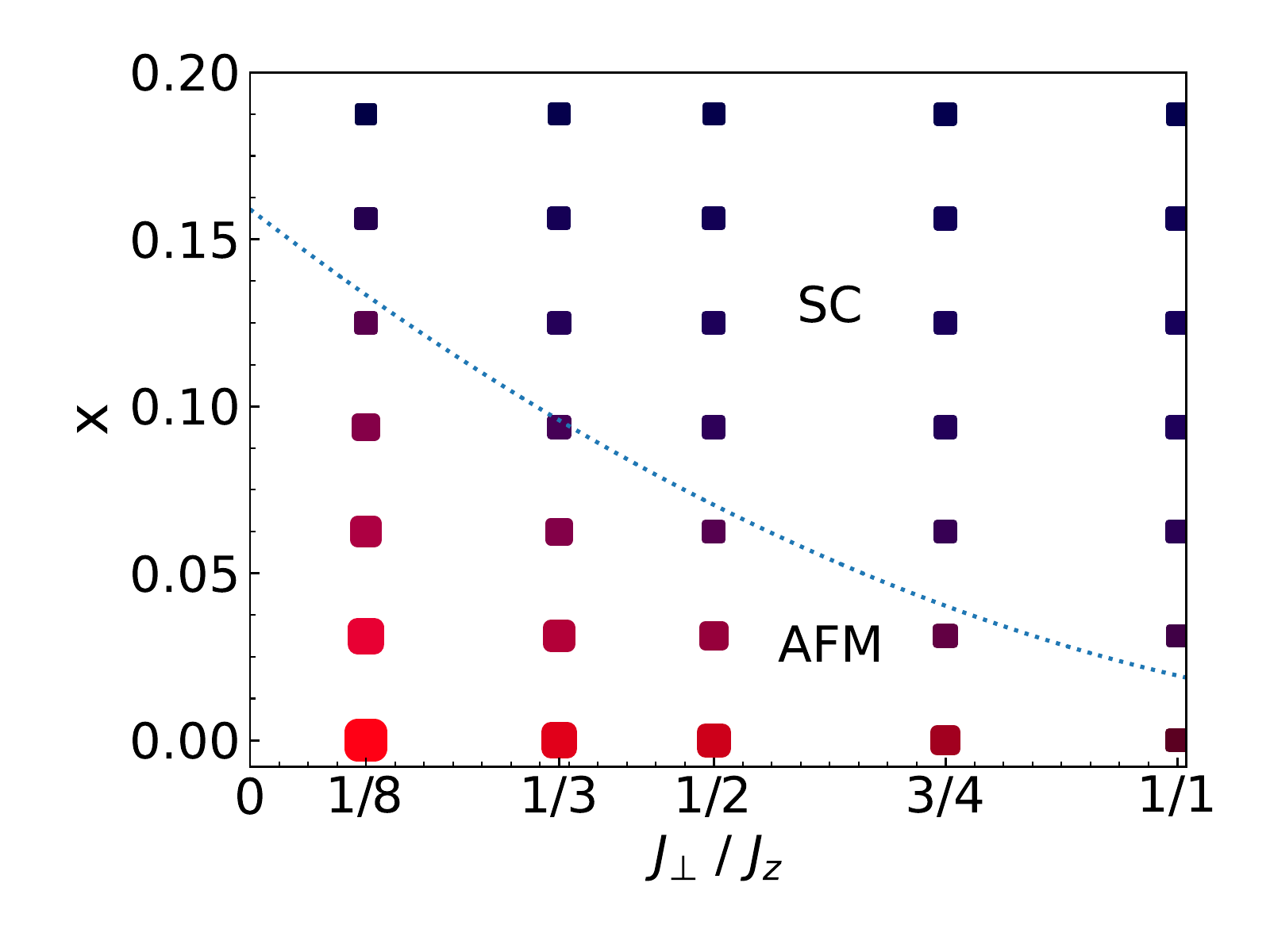}
\caption{An illustration of phase at $L=8$ and $U=1.25$. The size of symbols indicate the value of the correlation lengths
among SC and AFM. The color of symbols indicate the difference between them. The larger the AFM (SC) correlation length is, the more red (blue) the symbol is. The dotted line approximate indicates the boundary between them. The $x$ ticks are the value of $J_{\bot} / J_{z}$, $y$ ticks are the doping.}
\label{fig:pd1.25}
\end{figure}

To illustrate the competition between antiferromagnetism and superconductivity , we compare the correlation lengths of SC and AFM at different hole doping and $J_{\bot} / J_{z}$. The $J_{\bot} / J_{z}$ reflects the anisotropy of interlayer interaction, and when $J_{\bot} = J_{z}$  the interaction between two layers is isotropy Heisenberg interaction. First, we investigate the correlation lengths at different $J_{\bot}$ and $J_{z}$ but fixing their summation $U=-\frac{J_{z}}{4} - \frac{J_{\bot}}{2}$, and the results are shown in Fig.~\ref{fig:pd1.25}

\begin{figure}[h]
\includegraphics[width=0.45\textwidth]{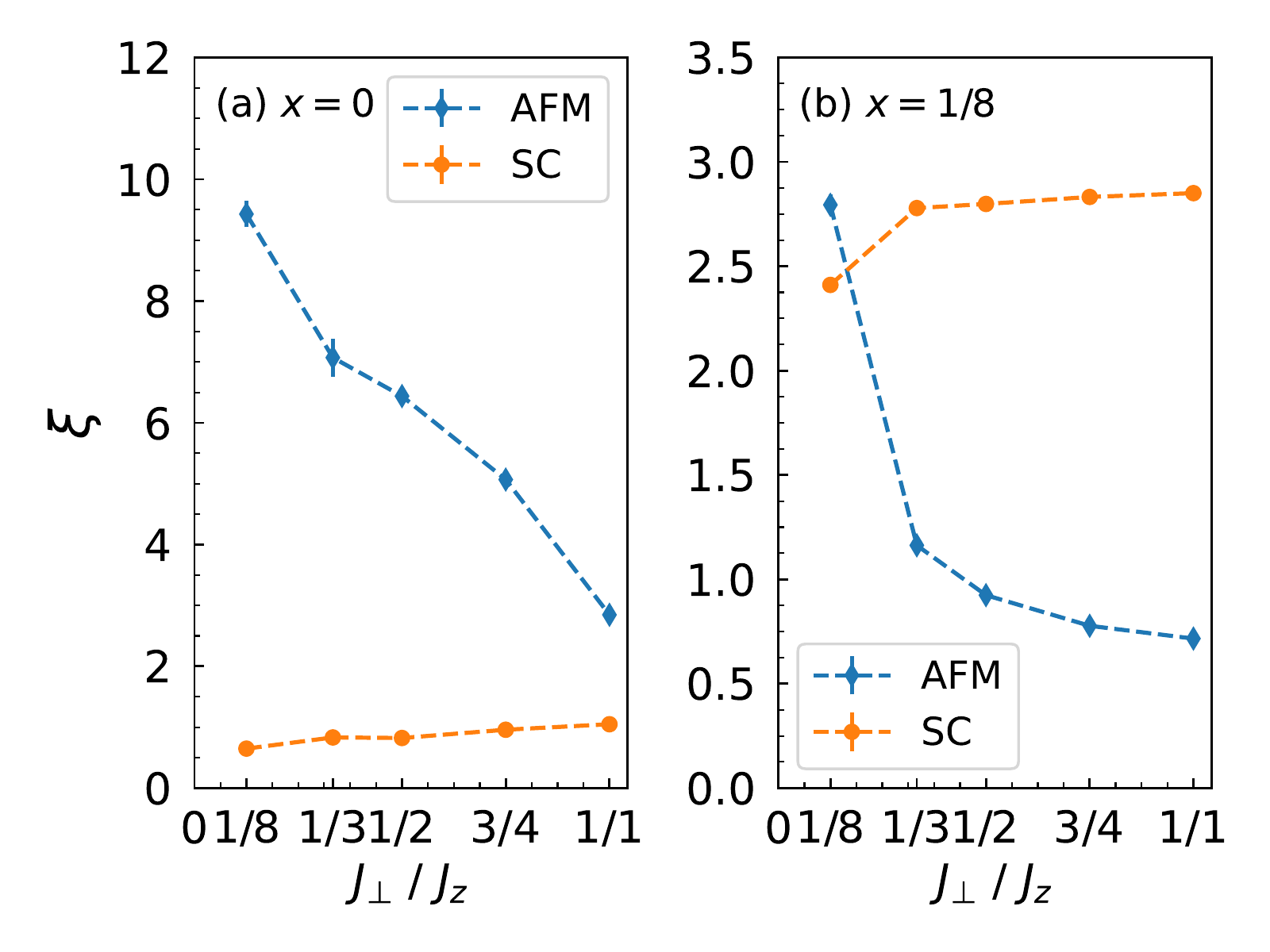}
\caption{Correlation lengths at $L=8$, $U=1.25$ at (a) half filling and (b) $x=\frac{1}{8}$ doping with different $J_{z}$ and $J_{\bot}$.
As $J_{\bot}/J_{z}$ increase, SC correlation length increase and AFM correlation length decrease, and AFM state is more sensitive to $J_{\bot} / J_{z}$ than SC.}
\label{fig:anisodep}
\end{figure}

This picture directly reveals the competition between antiferromagnetism and superconductivity. 
The advantage of this model is that one can enhance AFM or SC order by tuning the parameters
in Hamiltonian Eq.~\ref{eq:1}. The system favours SC order when $J_{\bot}$ is larger and favours AFM
order when $J_{z}$ is larger. An illustration of this can be seen at Fig.~\ref{fig:anisodep}.
By tuning these parameters, even at half filling the AFM order may be broken, and we will explain more latter.

\begin{figure}[h]
\includegraphics[width=0.45\textwidth]{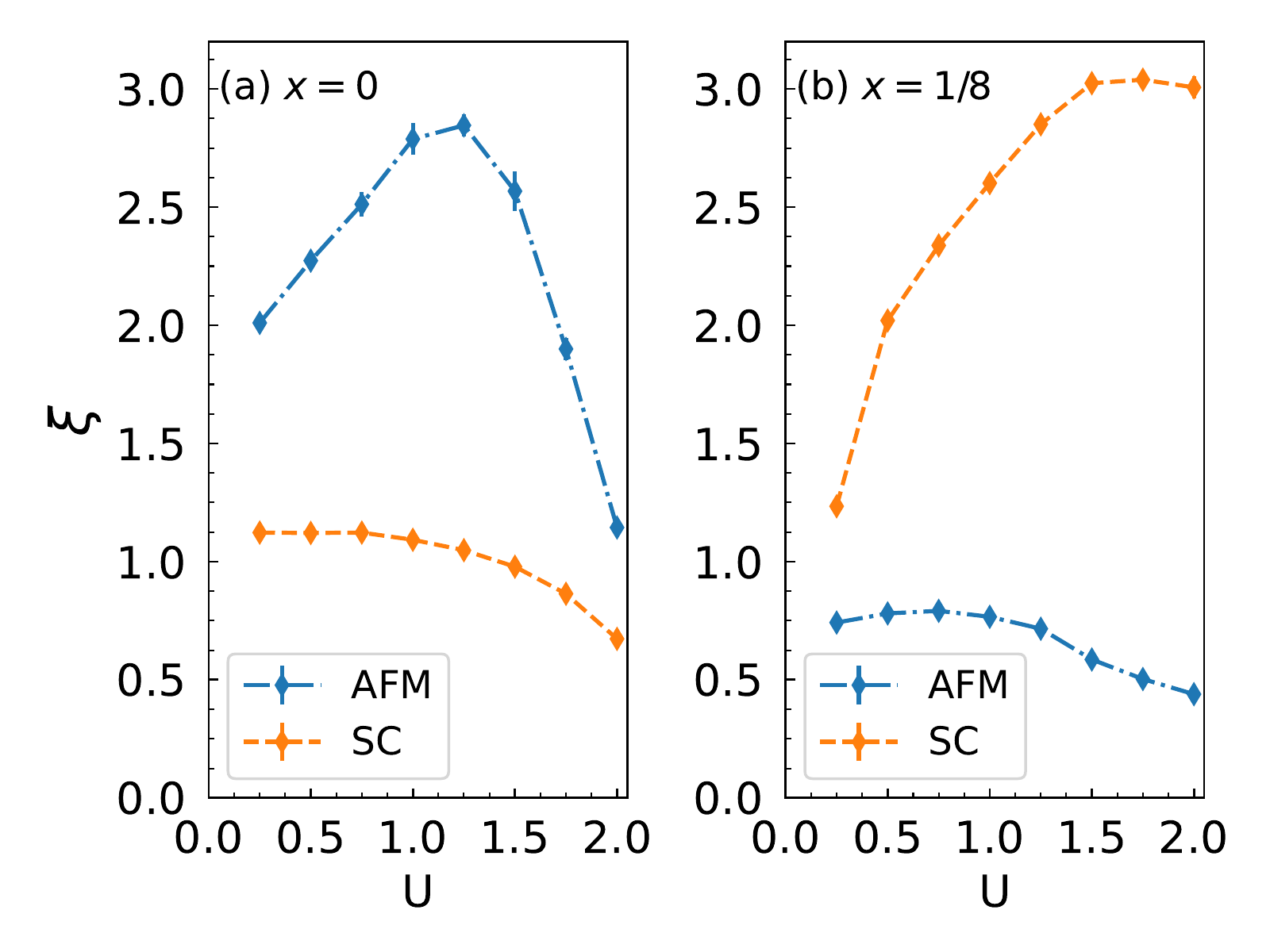}
\caption{Correlation lengths at $L=8$, $J_{z} = J_{\bot}$.
(a) At half filling, the AFM strength increases firstly and then decreases quickly, and the peak is around $U=1.25$.
The SC strength also drops around $U=1.0$.
(b) At $x=\frac{1}{8}$ for doping, the AFM strength decreases monotonically, and the SC strength increase monotonically.}
\label{fig:udep}
\end{figure}

Next, we enlarge $J_{z}$ and $J_{\bot}$ simultaneously, and keep the $J_{z} / J_{\bot} = 1$, for which the correlation length is shown at Fig.~\ref{fig:udep}.
The surprising thing is that at half filling, the AFM correlation length does not increases monotonically as $U$ increases.
It starts to decline when $U \approx 1.25$, and the SC correlation length also decrease.
At half filling, superconductivity does not emerge although the magnetism has been suppressed by tuning the value of $J_{\bot} / J_{z}$.
This implies that there maybe another order we have not discussed and the relationship between SC and AFM may be more complex at larger $U$ value.

\begin{figure}
\includegraphics[width=0.45\textwidth]{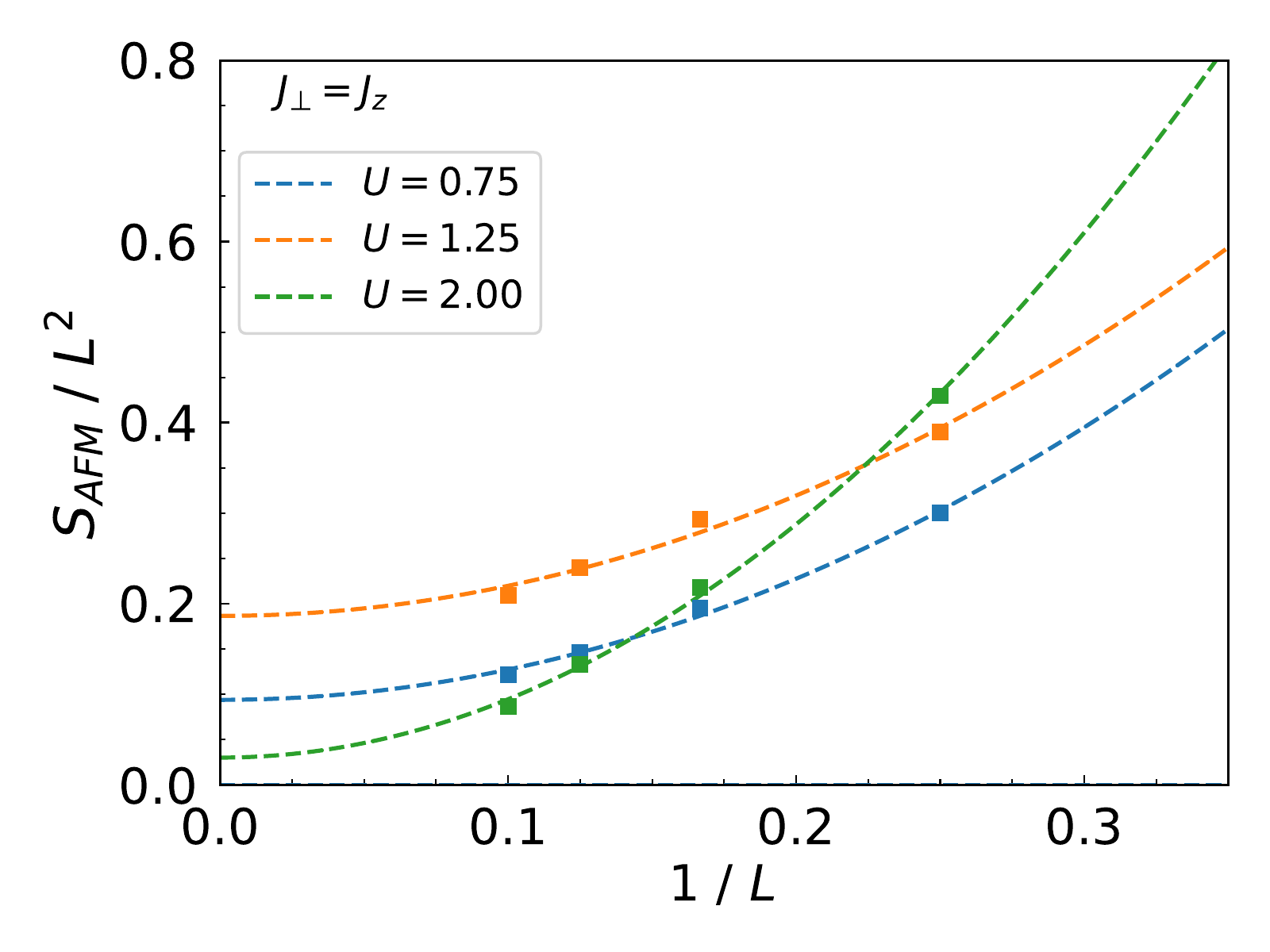}
\caption{Finite size scaling of $S_{AFM}$ at half filling for $J_{z} = J_{\bot}$.
One can see that the residual at thermodynamic of $U=2.00$ is much smaller than that of $U=1.25$.
This is consist with correlation lengths that we have shown In Fig.\ref{fig:anisodep}.}
\label{fig:fss1}
\end{figure}

This disappearance of AFM can also be confirmed by finite size scaling.
Base on the scaling hypothesis $S_{AFM}(L)/L^2 = a + b/L + c \xi^2 / L^2 $, we extrapolate the $S_{AFM}(L) / L^2$ to thermodynamic limit\cite{PhysRevX.3.031010}. Besides, the $b$ term can be ignored in AFM extrapolation since it comes from gapless excitations.
These results are shown in Fig.~\ref{fig:fss1}, where the long range AFM order developed at small $U$ and it starts to be broken as $U$ goes larger.

\begin{figure}
\includegraphics[width=0.45\textwidth]{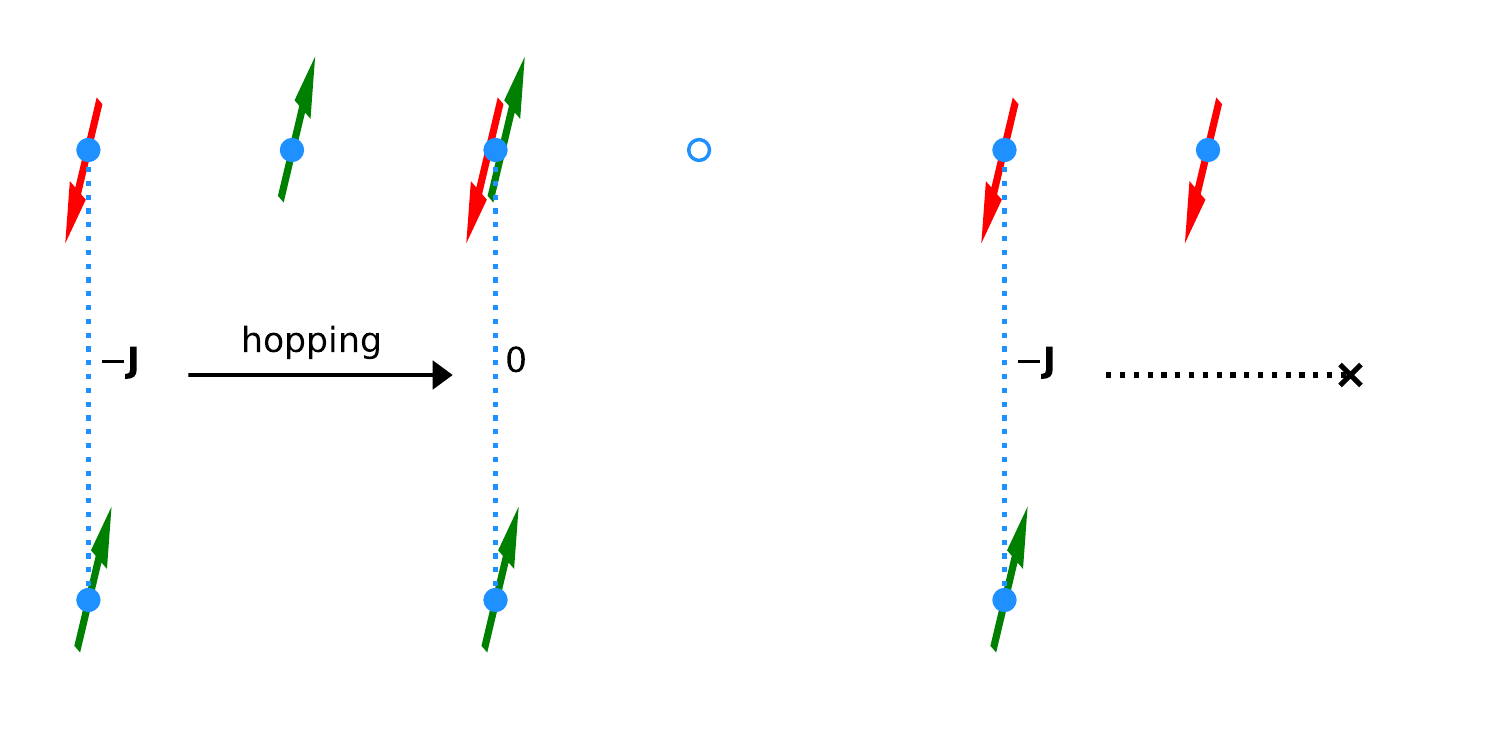}
\caption{A sketch of spin configuration shows that the interlayer interaction affects the intralayer order.}
\label{fig:p2}
\end{figure}

At first glance, the AFM decreases as $U$ increases, this may seem weird.
Recall the definition of $U$, $J_{\bot}$ and $J_{z}$, $U=-2g_1 - 4g_2$, $J_z = -8g_1$, $J_{\bot} = -8g_2$.
The interlayer Heisenberg interaction is proportional to Hubbard interaction $4U = J_{z} + 2J_{\bot}$.
When the interlayer interaction become stronger, the intralayer AFM order become more and more negligible.
As shown in Fig.~\ref{fig:p2}, the antiparallel spins allow a virtual hopping process,
and this process will make the energy of system become higher.
On the contrary, the parallel spins forbid this virtual hopping process, and may have lower energy.

\begin{figure}[h]
\includegraphics[width=0.45\textwidth]{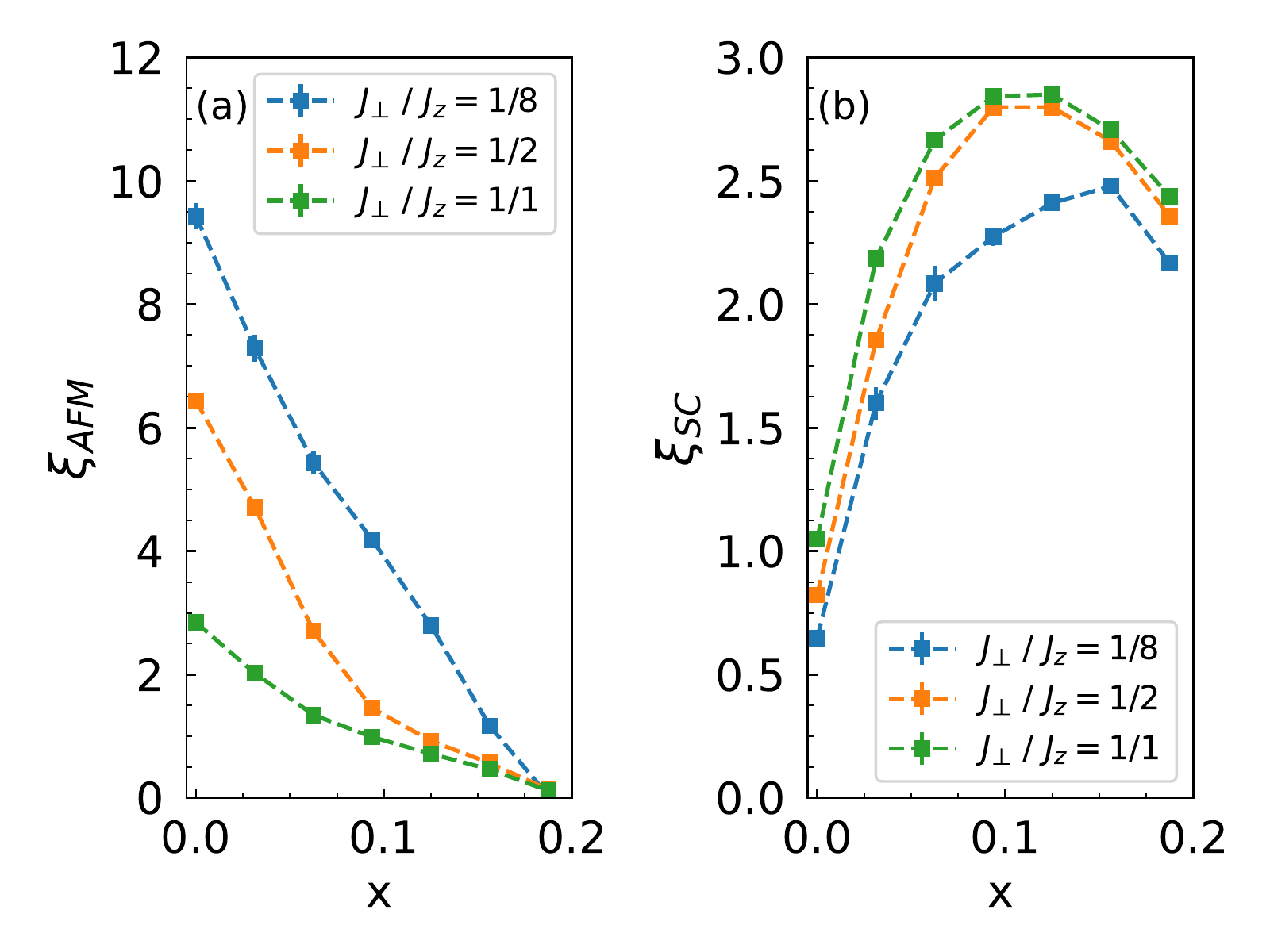}
\caption{AFM and SC correlation length at different electron number and $U$ is fixed at $1.25$.
(a) AFM orders prefer large $J_{z}$ and half filling, while (b) SC orders prefer large $J_{\bot}$ and finite doping.}
\label{fig:dopingdep}
\end{figure}

In Fig.~\ref{fig:udep}, one can see that the doping case is different from that at half filling.
The SC correlation length increases monotonically as $U$ increases, and AFM correlation length decreases.
At finite doping, the increasing interaction strength always suppress AFM order and favor SC order. Next, we show the doping dependence of AFM and SC order. From Fig.~\ref{fig:dopingdep},
one can see that the SC order prefers finite doping, and the optimal doping is depend on $J_{\bot} / J_{z}$.
At $J_{z} = J_{\bot}$, the optimal doping is around $1/10$.
It goes larger when $J_{z}$ increase, and as $J_{\bot}/J_{z} = 1/8$, it is around $1/6$.

\begin{figure}[h]
\includegraphics[width=0.45\textwidth]{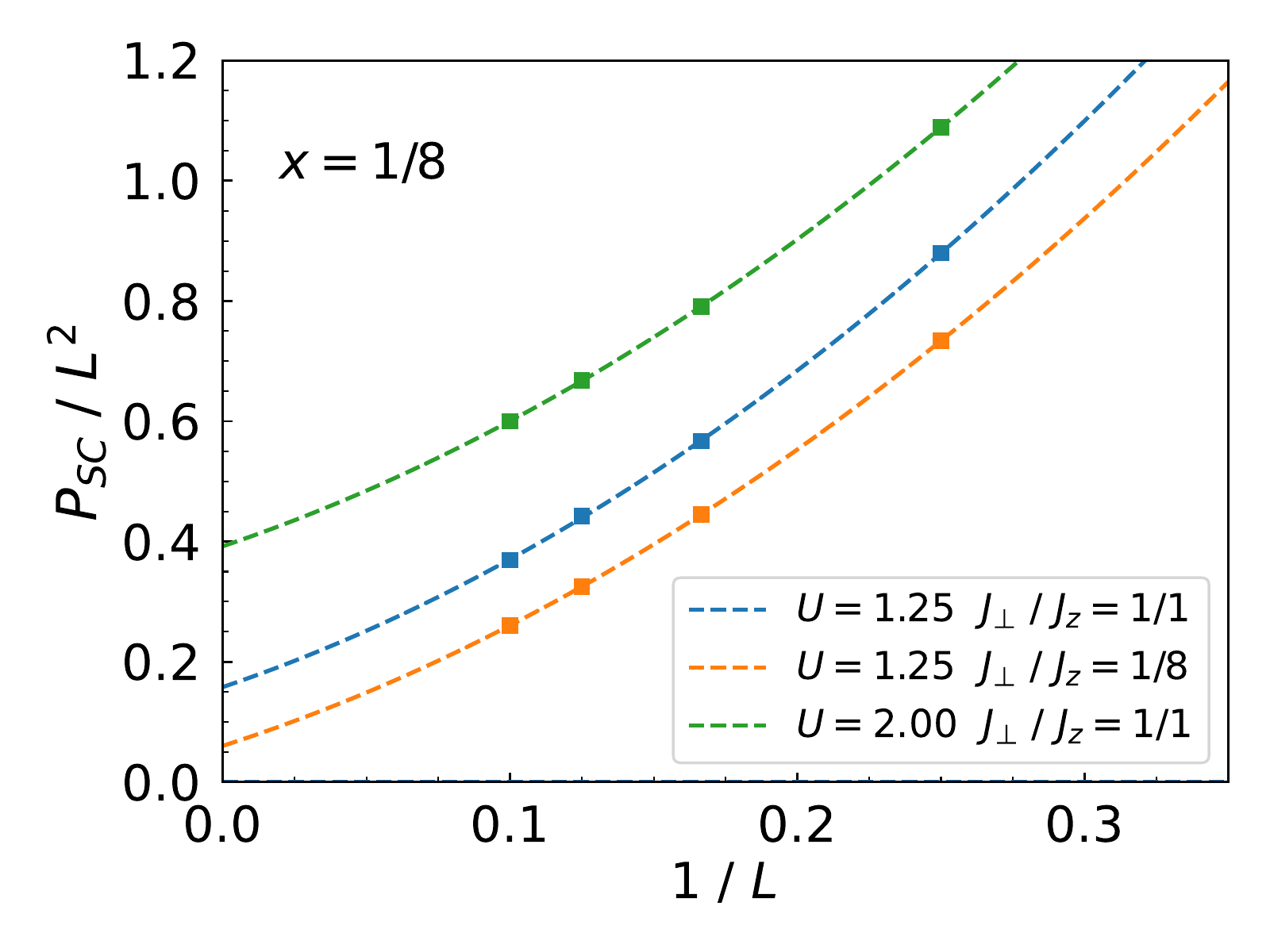}
\caption{Extrapolations of SC correlations at $L=8$ and $x=\frac{1}{8}$. The SC strength become larger when either
enlarge $U$ or enlarge $J_{\bot}$. This is consist with correlation lengths we compute before.}
\label{fig:fss2}
\end{figure}

One may notice that the SC correlation lengths are smaller than AFM correlation lengths,
and may wonder whether there is long range SC order or not.
To characterize the long range properties of SC order, we also extrapolate SC structure factor to thermodynamic limit in Fig.~\ref{fig:fss2}.
At $L=10$ or $L=6$, $\frac{1}{8}$ doping correspond to hole number $25$ and $9$, these are not close shell fillings and we average the SC correlations of the nearest close shell fillings around them.
From the finite size scaling results, one can see that the SC long range order shall be established at proper choice of parameters.
These results confirm that larger $U$ and $J_{\bot}$ favours SC order at finite doping.

\begin{figure}[h]
\includegraphics[width=0.45\textwidth]{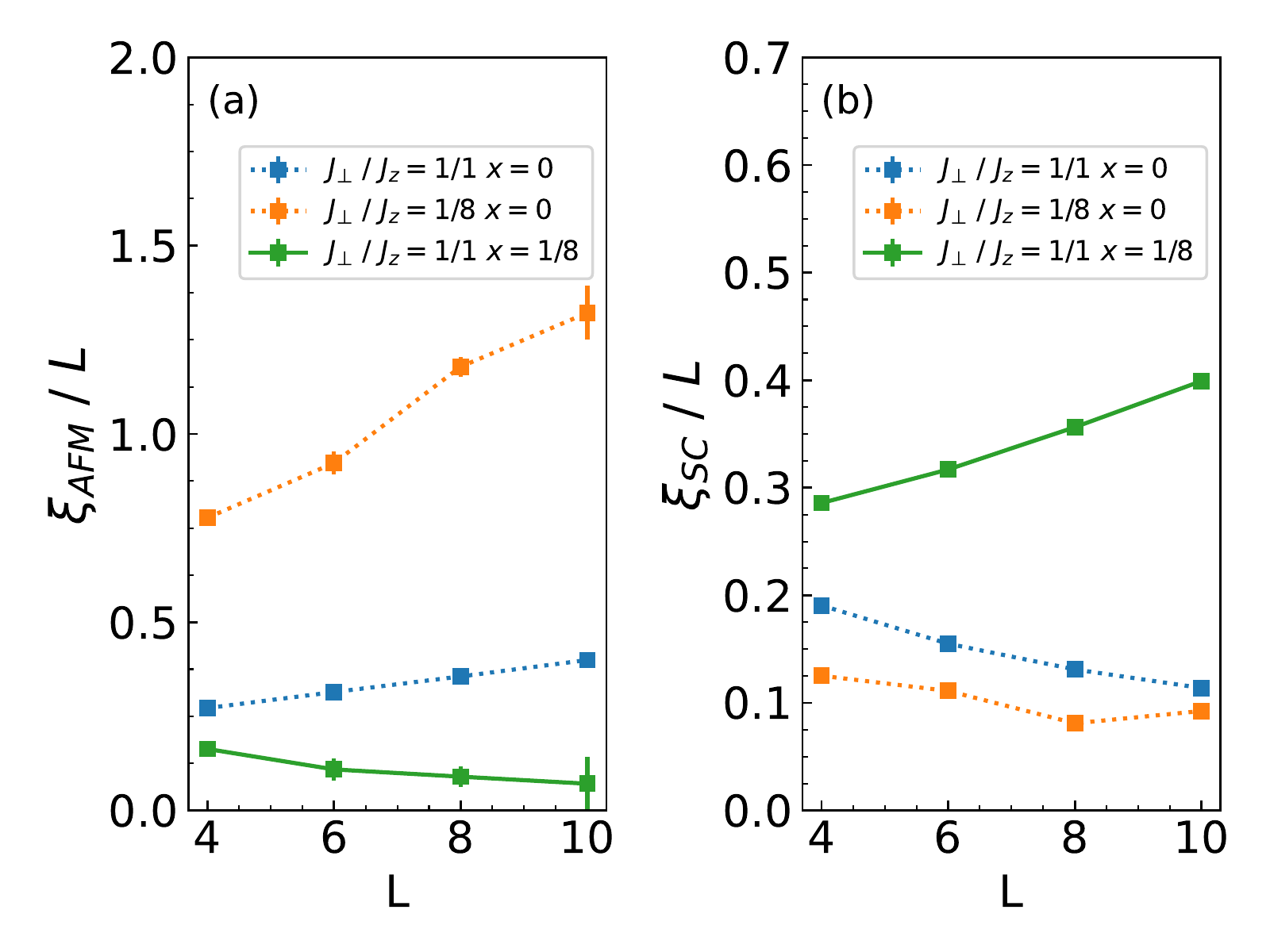}
\caption{Correlation lengths at different lattice size, (a) is AFM and (b) is SC.
In this picture, dotted lines are at half filling, and the solid lines are at $1/8$ doping and $U = 1.25$ is fixed.
Where we can see the results of $L=8$ can characterize the properties of system qualitatively, so we may use those results to reveal the competition between antiferromagnetism and superconductivity.
}
\label{fig:versize}
\end{figure}

Finally, we check the effects of system size.
Most of our results are simulated on lattice with $L=8$, and in that case, there are $2\times8^2$ sites in total, which is fairly large.
We choose this relative large lattice size because of the special form of interaction in Hamiltonian Eq.~\ref{eq:1}.
In our PQMC simulations, the computational cost is nearly the same as simulations on $4\times8^2$ sites of ordinary Hubbard model. In Fig.~\ref{fig:versize}, we show the SC and AFM correlation lengths at different lattice size.
One can see that, in some sense, $L=8$ is not large enough, when finite size scaling results assert that
there is AFM or SC long range order, the corresponding $\xi / L$ is increasing and the other $\xi / L$ is decreasing.
So we can not use $\xi$ at $L=8$ to assert whether there is a long range order or not.
Fortunately, the results at different $L$ are qualitatively consist with each others, so we can use the correlation lengths
$\xi$ to compare the strength of SC and AFM. For analysis of long range order, we still resort to finite size scaling.

\section{Summary}

In this work, we utilize a sign problem free model to investigate the competition between SC and AFM order.
By performing PQMC simulations, we compare the correlation lengths at different parameters, and use finite size scaling technique to study the long range behaviours.
Our results show that at doping case, antiferromagnetism is suppressed by $J_{\bot}$ interaction, and the superconductivity shall be enhanced. At half filling, the superconductivity will not emerge although the antiferromagnetism will be suppressed by enlarging the interaction strength. The antiferromagnetism does not increase or decrease monotonically with interaction strength $U$, and it has a peak around $U = 1.25$. The optimal doping of SC is depend on the $J_{\bot}/J_{z}$, and the optimal doping is a little larger when $J_{z}$ dominate. The finite size scaling results are qualitatively consist with the correlation lengths.
Our results may provide some new aspects of understanding superconductivity and its parents materials, and may also simulate further cold atom experiments to realize such model to tune the competition between SC and AFM order in one system.
\noindent
\underline{\it Acknowledgments} ---
This work was supported by the NSFC (No. 11974049). The numerical simulations were performed at the HSCC of Beijing Normal University and on Tianhe-2JK in the Beijing Computational Science Research Center.
\bibliography{ref}
\end{document}